# Investigation of the effect of the grain sizes on the dynamic strength of the fine-grained alumina ceramics obtained by Spark Plasma Sintering

Melekhin N.V.[1], Boldin M.S.[1(*)], Bragov A.M.[1], Filippov A.R.[1], Popov A.A.[1], Shotin S.V.[1], Nokhrin A.V.[1], Chuvil'deev V.N.[1], Murashov A.A.[1], Tabachkova N.Yu.[2,3]

[1] Lobachevsky State University of Nizhny Novgorod, Nizhny Novgorod, Russia

[2] National University of Science and Technology "MISIS", Moscow, Russia

[3] A.N. Prokhorov General Physics Institute, Russian Academy of Science, Moscow, Russia

boldin@nifti.unn.ru

**Abstract**

The results of dynamic strength tests of the alumina ceramics with various grain sizes are presented. The ceramics were obtained by Spark Plasma Sintering (SPS) of industrial submicron and fine α-$Al_2O_3$ powders. The heating up was performed with the rate of 10 ºC/min; the grain sizes in the ceramics was controlled by varying the SPS temperature and the heating rate as well as by varying the initial sizes of the α-$Al_2O_3$ particles in the powders. The ceramics had a high density (over 98%) and a uniform fine-grained microstructure (the mean grain sizes varied from 0.8 to 13.4 μm). The dynamic compressing tests were carried out by modified Kolsky method with using split Hopkinson pressure bar. The tests were performed at room temperature using a 20-mm PG-20 gas gun with the strain rate of ~$10^3$ $s^{-1}$. The dependence of the dynamic ultimate strength of alumina on the grain size was found for the first time to have a non-monotonous character (with a maximum). The maximum value of the dynamic ultimate compression strength ($\sigma_Y$ = 1060 MPa) was provided at the mean grain size of ~2.9-3 μm. The reduction of $\sigma_Y$ for alumina in the range of submicron grain sizes was shown to originate from the reduction of the relative density of the ceramics sintered at lower SPS temperatures.

---

(*) corresponding author (boldin@nifti.unn.ru)

The reduction of $\sigma_Y$ of the ceramics in the range of the micron-scale grain sizes was suggested to be caused by formation of internal strain near the grain boundaries. The hardness $H_v$ of alumina was found to decrease and the minimum fracture toughness (according to Palmquist) $K_{IC}$ – to increase with increasing grain size. The results of the present study demonstrated the absence of the necessity to provide forming the nanostructure in the ceramics to ensure high dynamic strength characteristics.

**Introduction**

Alumina is one of the most common ceramic materials applied widely in general and sophisticated mechanical engineering and in power engineering [1-3]. Its wide application originates from a good combination of hardness, fracture toughness, high melting point, and low cost.

A low dynamic strength of the $Al_2O_3$-based ceramics is one of the key problems in the way to the expansion of the field of application for the ones [2, 4-6]. This limits essentially the field of application of the $Al_2O_3$ ceramics in the products subjected to various impact loads in the course of operation. Forming a uniform fine-grained (FG) microstructure in the ceramics is a promising way to improve the mechanical properties of alumina [1, 3, 4, 7, 8]. Some researchers stated a problem of forming a structure with the ultimately small grain sizes in the ceramics to achieve the best mechanical properties [7, 8]. Note also that the dynamic strength of nano- and fine-grained alumina-based ceramics is studied extensively at present [5, 6].

However, there are no systematic data on the effect of the grain sizes on the dynamic strength of alumina in the literature. It is related to the technological limitations of conventional sintering methods, which don't allow providing a high density and varying the grain sizes in binderless alumina in a wide range – from the submicron level to several tens microns simultaneously.

Spark Plasma Sintering (SPS) is one of efficient methods for obtaining the FG alumina-based ceramics [9-11]. In SPS, a graphite mold with a powder inside is heated by passing the millisecond high-power electric current pulses [9-11]. SPS is performed in vacuum or in an inert ambient with simultaneous application of a uniaxial pressure. SPS is featured by the capabilities to control main sintering parameters (the heating rate, the holding time and temperature, the pressure, etc.) directly in the course of sintering and to perform the stepwise sintering regimes [12, 13]. This provides wide capabilities of SPS in controlling the microstructure parameters of the ceramics as well as allows forming a fine-grained structure with improved mechanical properties in the ceramics [9-13].

The present study was aimed at the investigation of the effect of the grain sizes on the dynamic strength of the fine-grained alumina.

**Materials and methods**

The ceramic alumina specimens obtained by SPS were the objects of investigations. The submicron $\alpha$-$Al_2O_3$ powders from Taimei Chemicals Co. Ltd. (Series #1) and the fine ones from Alfa Aesar - A Johnson Matthey Co. (Series #2) were used as the initial raw materials. According to the vendors' specification, the powder of Series #1 had the mean initial particle sizes of ~0.2 μm, the one of Series #2 — near 1 μm. There were some amorphous layers on the surfaces of the submicron particles of the powders of Series #1.

The specimens of 12 mm in diameter were compacted using Dr. Sinter® model SPS-625 setup. The sintering was performed in vacuum; the uniaxial pressure applied was 70 MPa. A two-stage heating regime was applied: heating with the rate of 100 °C/min up to 600 °C and slow heating with the rate of 10 °C/min up to the temperature of the finish of shrinkage (Fig. 1a). The temperature on the graphite mold surface (T1) was measured using IR-AHS2 pyrometer. Based on the investigations carried out earlier and of the comparison of the readouts of the pyrometer (T1 in Fig. 1a) and of the reference thermocouple mounted onto the specimen surface, the values of T1 in the temperature range over 600 ºC were recalculated in the actual temperature of the specimen (T2) using an empirical relation: $T2 = 1.1686 T1 - 43.416$. The effective shrinkage of the specimen ($L_{eff}$) was measured using the built in dilatometer of Dr. Sinter® model SPS-625 setup. The actual shrinkage of the powder (L) was calculated by subtracting the contribution of the thermal expansion of the setup – mold – specimen system ($L_0$) according to [14] (Fig. 1b).

After sintering, the specimens were polished up to the roughness level of 5 μm or less.

The microstructure of the ceramics was studied using Jeol® JSM-6490 SEM. The mean grain sizes were determined by the secant method using GoodGrains software.

The density of the specimens ($\rho$) was measured by hydrostatic weighting. The theoretical density of $Al_2O_3$ was taken $\rho_{th} = 4.05$ g/cm$^3$.

The Vickers hardness ($H_V$) was measured using Struers® Duramin™ 2 microhardness tester with the load of 2 kg. The minimum fracture toughness coefficient $K_{IC}$ was calculated by Palmquist method from the length of the longest radial crack. In the calculation of $K_{IC}$, the magnitude of the elastic modulus was taken $E = 350$ GPa. The uncertainties of determining $H_v$ and $K_{IC}$ were ±0.5 GPa and ±0.3 MPa·m$^{1/2}$, respectively.

The dynamic tests of the ceramics were carried out using the modified Kolsky method with the split Hopkinson pressure bar at the strain rates of ~ $10^3$ s$^{-1}$. For testing the specimens of 12 mm in diameter and 6 mm in height, a 20 mm gas gun RSG-20 was used. The tests were carried out at room temperature. The measuring rods (the loading rod and the support one) of 20 mm in diameter and 1 m long for the compression tests were made from high-strength steel with the yield strength 1600 MPa. The magnitude of the dynamic ultimate strength ($\sigma_Y$) was determined from the value of the maximum stress. The test procedure was described in details in [15]. Note also that this procedure is used extensively worldwide to evaluate the dynamic strength of the fine-grained alumina-based ceramics [16, 17].

**Experimental results**

The dependencies of the temperature and pressure on the SPS time and corresponding temperature curves of the shrinkage L for the alumina powders from Series #2 are presented in Figs. 1a and b, respectively as examples.

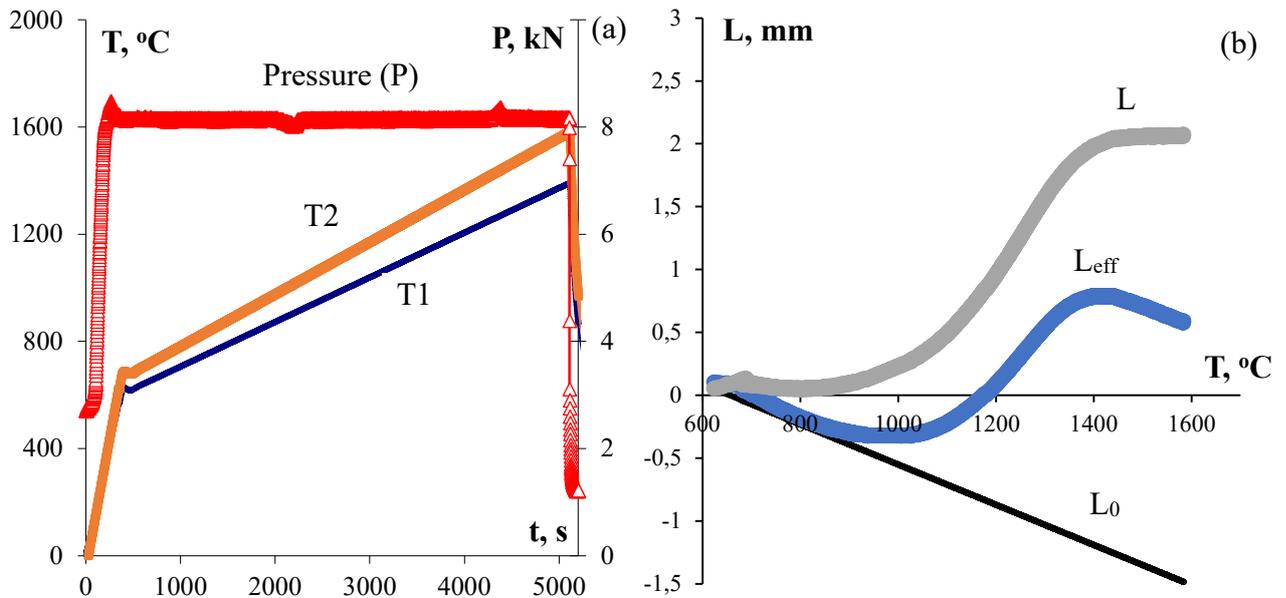

Figure 1 – SPS diagrams for the ceramics: (a) dependencies of temperature (T) and pressure (P) on the heating time; (b) temperature curves of shrinkage for the powders from Series #2

To make the ceramic specimens, a low-rate heating (10 ºC/min) was used. It allowed avoiding the cracking of the ceramics. The results of the metallographic and Electron Microscopy investigations evidenced the absence of the microcracks and of other defects. The low-rate heating up allowed also minimizing the thermal field nonuniformity inside the ceramic specimens [18, 19] and, hence, avoiding the nonumiformities of the microstructure and properties.

The temperature curves of the shrinkage L(T) had a conventional three-stage character [20]. It is worth noting the third high-temperature stage expressed clearly in the dependence L(T).

At this stage, the density of ceramics is already close to its theoretical value ($\rho \sim \rho_{th}$) and an intensive grain growth begins in the ceramics [20]. It allows controlling the mean grain sizes in the ceramics by choosing the temperature and/or the duration of Stage III.

As a result of sintering, the ceramic specimens with high relative density (> 99% for the powders from Series #1 and > 98% for the ones from Series #2) were obtained. The relative densities of the specimens sintered from the submicron powder Series #1 were 1-1.5% greater than the ones of the specimens sintered from powder Series #2 with more coarse grains (Table 1). This result agrees well with the published data on the accelerated powder sintering at reduced dispersion level [20].

The investigations of the microstructure of ceramic specimens by SEM are presented in Fig. 2 and in Table 1.

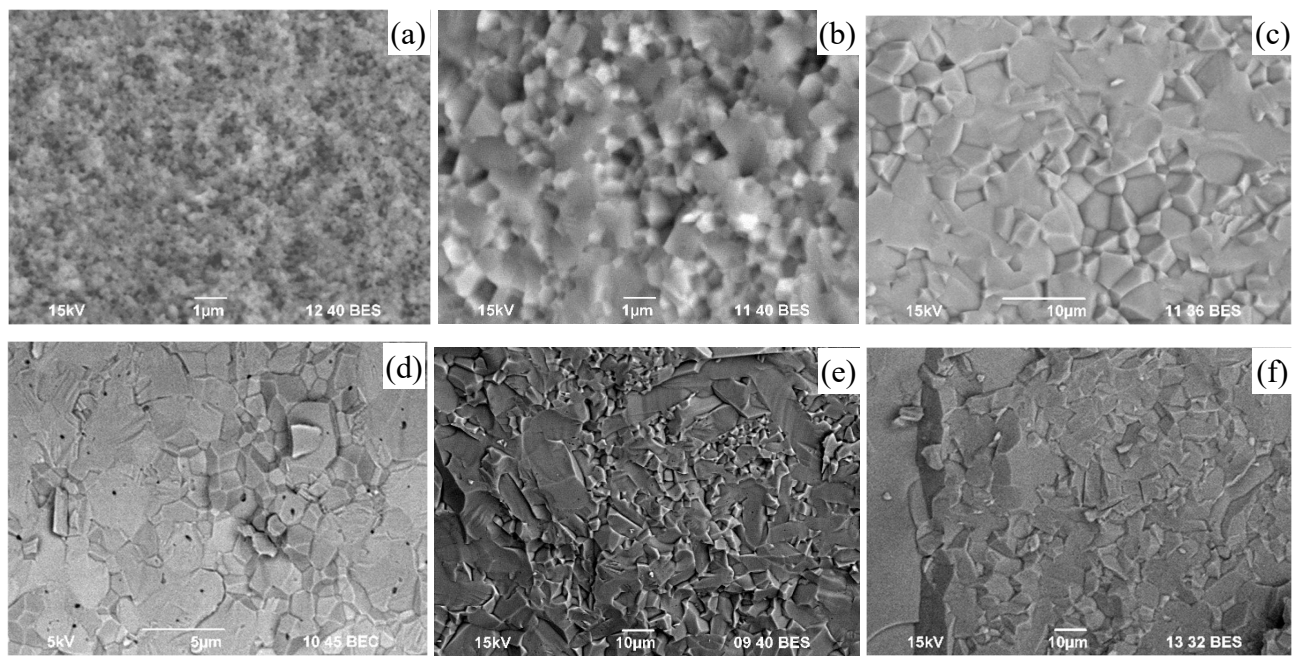

Figure 2 – Microstructure of alumina obtained by SPS of the submicron (a, b, c) and fine (d, e, f) powders at 1320 (a), 1420 (b), 1470 (d), 1520 (c), 1530 (e), and 1600 °C (f). SEM of the fractures

Table 1 – Results of microstructure investigations and mechanical properties of alumina obtained at various SPS temperatures

| Powder | SPS mode | | Microstructure parameters and ceramic properties | | | | |
|---|---|---|---|---|---|---|---|
| | $V_h$, °C/min | $T$, °C | $\rho/\rho_{th}$, % | $d$, μm | $H_v$ GPa | $K_{IC}$, MPa·m$^{1/2}$ | $\sigma_Y$, MPa |
| 1 | 10 | 1320 | 99.46 | 0.5-0.8 | 22.3 | 2.2 | 895 |
| 1 | 10 | 1420 | 99.66 | 1.6-1.8 | 20.0 | 2.4 | 940 |
| 1 | 10 | 1520 | 99.72 | 5.1-5.6 | 18.6 | 2.5 | 990 |
| 2 | 10 | 1470 | 97.98 | 2.1 | 18.4 | 2.7 | 1060 |
| 2 | 10 | 1530 | 98.24 | 12.9 | 16.2 | 2.0 | 925 |
| 2 | 10 | 1600 | 98.25 | 20.0 | 16.1 | 1.7 | 900 |

The ceramics had uniform fine-grained microstructure, no traces of the abnormal grain growth were observed (Fig. 2). The microstructure of the specimens was uniform in the longitudinal section as well as in the cross-section. The large pores inside the grains or at the grain boundaries were absent.

The increasing of the SPS temperature for the submicron powder #1 from 1320 up to 1520 °C resulted in an increase in the mean grain sizes in the ceramics from ~0.5-0.8 μm up to 5.1-5.6 μm. The increasing of the sintering temperature for the powder #2 from 1470 up to 1600 °C resulted in an increase in the mean grain sizes from 2.1 up to 20 μm (Table 1). The increasing of the sintering temperature in 200-230 °C resulted in an insufficient increase in the relative density of the ceramics in 0.26-0.27%.

Fig. 3a presents the dependencies of the microhardness $H_v$ and of the minimal fracture toughness coefficient $K_{IC}$ on the grain size. As one can see from the plot presented in Fig. 3a, the increasing of the grain size in the ceramics leads to a drastic reduction of the microhardness. The ceramic with the mean grain size of ~0.5-0.8 μm had the microhardness of 22.3 GPa. The increasing of the grain size up to 5.1 μm and up to 20 μm resulted in a decrease in the ceramic microhardness down to 18.6 and 16.1 GPa, respectively.

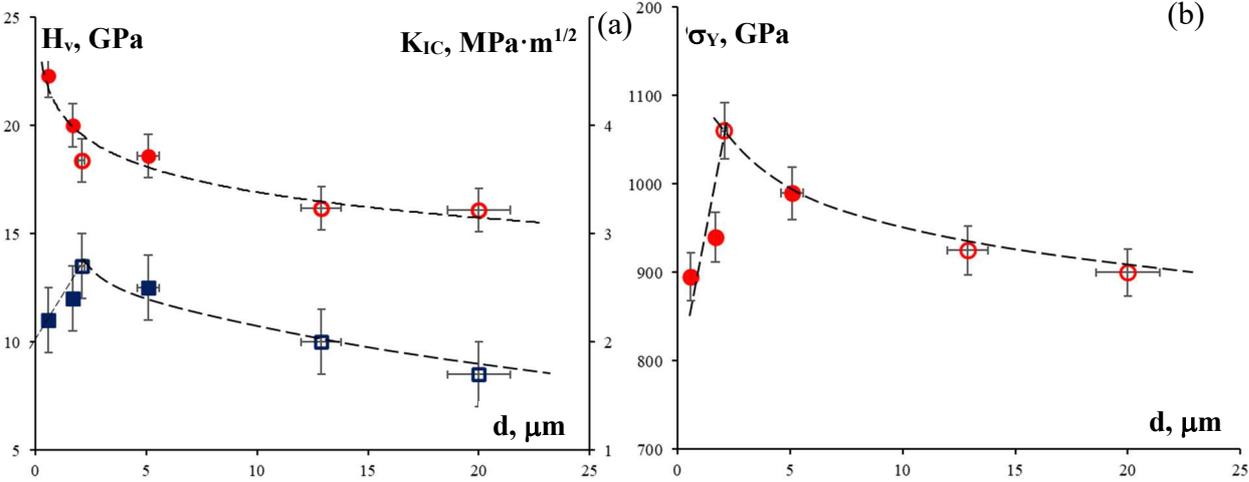

Figure 3 – Dependencies of the mechanical properties of alumina on the grain size: (a) $H_v$ (circles) and $K_{IC}$ (squares); (b) $\sigma_Y$. Filled markers – ceramics made from powder #1, empty markers – from powder #2

Fig. 4 presents typical stress – strain curve obtained from the dynamic testing of the alumina specimens. Fig. 3b and Table 1 summarize the results of investigations of the effect of the grain size on the dynamic ultimate strength of alumina.

The curves $\sigma(\varepsilon)$ had typical form for the curves in the case of dynamic compression of the ceramic specimens (see [16, 17]). During the first pulse, the fracture of the specimen as a whole takes place. Then, at further compression during the next pulses, the fragments are compressed. Since the cloud of fragments has a different area from the one of the initial specimen, the values of the dynamic strength appeared to be overestimated.

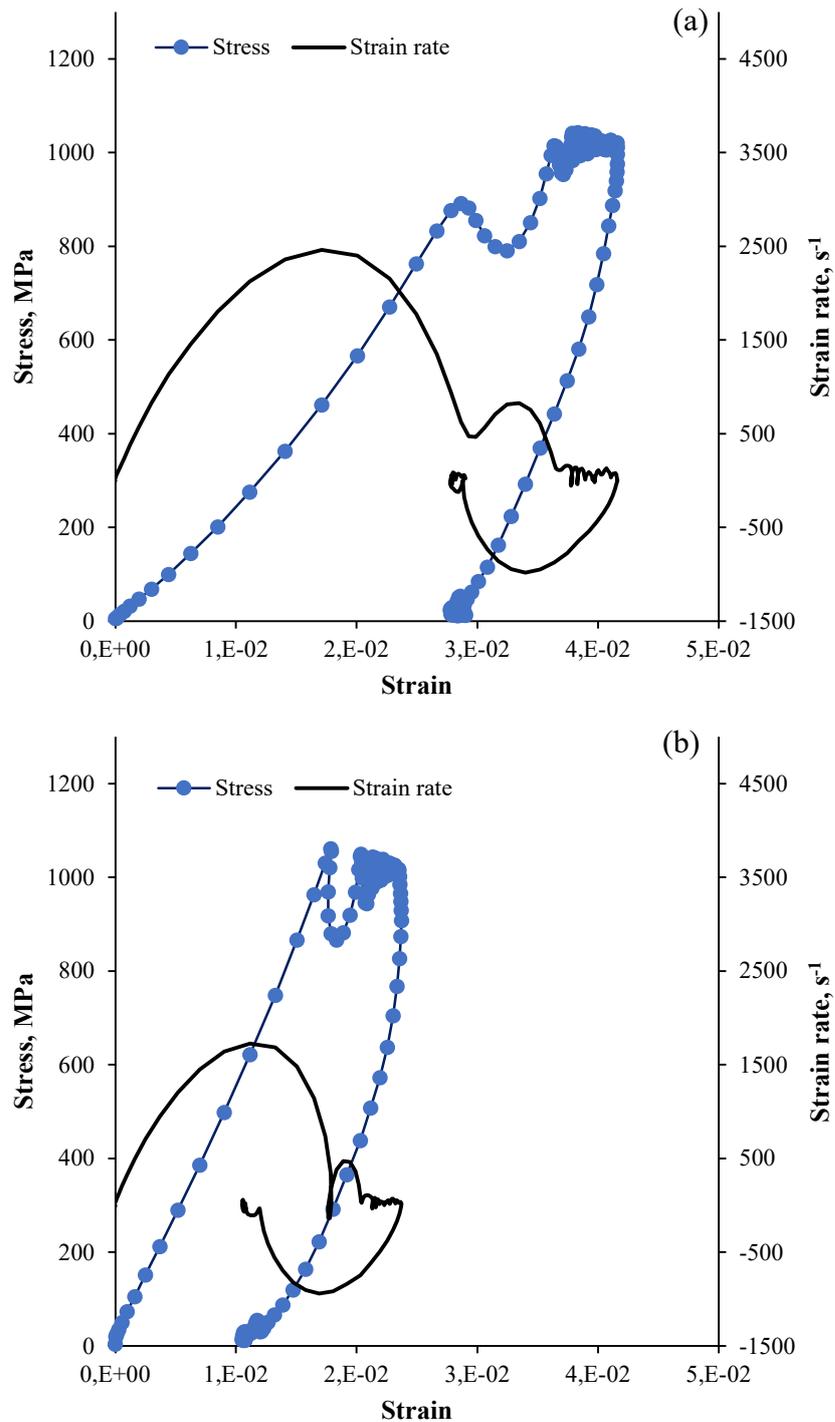

Figure 4 – Typical stress-strain curves of the alumina specimens with the ultimate strength $\sigma_Y = 895$ MPa (a) and 1060 MPa (b)

The analysis of the results presented in Table 1 shows the dependence of the dynamic ultimate strength on the mean grain size to have a nonmonotonous character with a maximum corresponding to the grain sizes ~2.9-3 μm. The increasing of the grain sizes from 0.5-0.8 μm up to 2.9- 3 µm resulted in an increase in the ultimate dynamic strength of alumina from 890 MPa up to 1060 MPa. Further increasing of the ceramic grain sizes up to 5.6 µm resulted in a decrease in the ultimate dynamic strength down to 990 MPa.

**Discussion**

Let us analyze the effect of the grain sizes on the mechanical properties of the fine-grained alumina.

The analysis of the results of the hardness measurements shows $H_v$ to decrease monotonously from 22.3 GPa down to 16.1 GPa with increasing mean grain size (d) from 0.5-0.8 µm up to 20 µm. The dependence $H_v(d)$ can be interpolated by the dependence $H_v - d^{-1/2}$ with a good precision (Fig. 5). Such a type of dependence is used often to describe the effect of the grain size on the hardness of the ceramics [21], although it is worth noting that the origin of the linear character of the dependence $H_v - d^{-1/2}$ in the metals and in the ceramics are quite different (see [21, 22]).

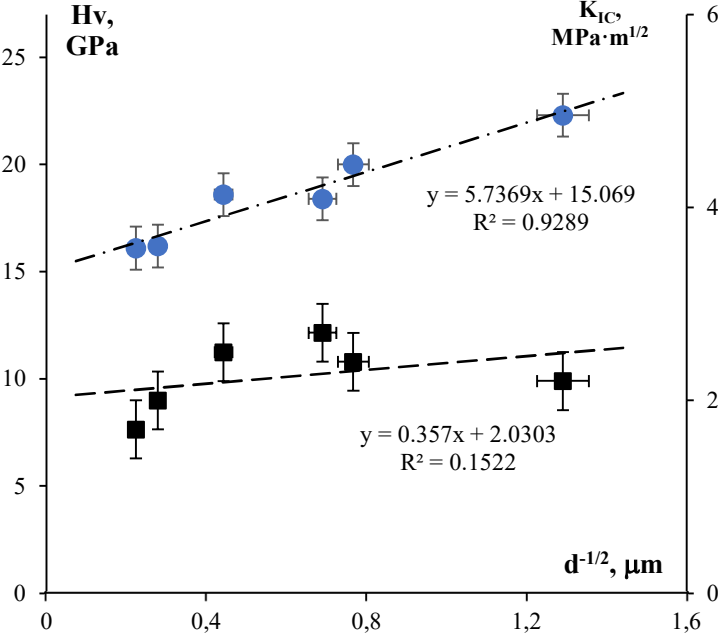

Figure 5 – Dependence of the microhardness (large markers) and of the fracture toughness (squares) on the grain size in the $H_v - d^{-1/2}$ and $K_{IC} - d^{-1/2}$ axes

No essential effect of the grain sizes on the minimal fracture toughness coefficient was observed. The maximum scale of the fracture toughness coefficient variation $\Delta K_{IC} = 0.5$ MPa·m$^{1/2}$ (from 2.2 to 2.7 MPa·m$^{1/2}$) exceeded the scatter of the properties of a specimen (± 0.3 MPa·m$^{1/2}$) only slightly. The confidence coefficient of linear approximation of the dependence $K_{IC} - d^{-1/2}$ was very small (Fig. 5).

The dependence of the ultimate dynamic strength on the grain sizes $\sigma_Y(d)$ had well-expressed two-stage character. The reduction of $\sigma_Y$ with increasing grain size from 2.1 to 20 μm is an expected result since the grain growth leads to a reduction of the ceramic strength characteristics usually [1, 3, 7, 8, 23].

The decreasing of $\sigma_Y$ in the range of small grain sizes (0.5 – 2 μm) was quite an unexpected result – as it has been already noted in Introduction, the formation of the fine-grained microstructure is a well-known method of improving the mechanical properties of the ceramics. One can suggest the decreasing of $\sigma_Y$ in the ultrafine-grained (UFG) alumina to originate from the decrease in the density of the ceramics – as one can see in Table 1, the ceramics with small grain sizes were sintered at reduced temperatures. The reduction of the SPS temperature for powders #1 from 1520 down to 1320 °C resulted in a decrease in the relative density from 99.72% down to 99.46%.

The nano- and submicron-sized pores formed at reduced SPS temperatures may be the centers of nucleation and further rapid propagation of the brittle cracks during the dynamic loading of the ceramics.

The presence of the amorphous layers on the surfaces of initial submicron powders (Fig. 6a) can be the second factor promoting the reduction of the dynamic strength of the ceramics with small grain sizes. Note, that free volume is the good criterion for evaluation of the state of the grain boundary for the ceramics [24-26].

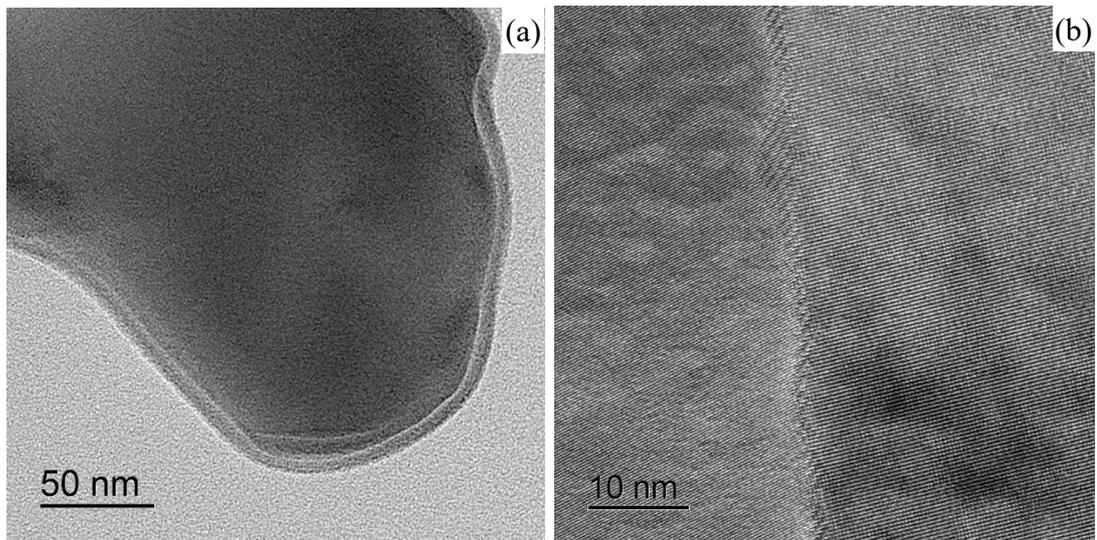

Figure 6 – Fragment of the microstructure of initial powder from Series #1 (a) and of the grain boundary of alumina sintered from this powder by SPS (b). TEM

The results of high-resolution TEM investigations demonstrated the grain boundaries in the sintered UFG ceramics to have the crystal structure typical for the grain boundaries (Fig. 6b). No amorphous structure inclusions were observed at the grain boundaries. It allows concluding the crystallization of the amorphous phase, which transformed into the grain boundaries of the UFG ceramics to take place in the course of SPS. The amorphous phase is known to contain an excess free volume [27, 28]. One can expect the transformation of the amorphous phase into the crystalline one in the course of rapid heating and sintering at reduced temperatures to result in the appearance of an excess density of the dislocation or vacancy-type defects at the grain boundaries in the UFG alumina. Because of a small duration of the sintering process and of low grain growth intensity, the defects arising at the grain boundaries have not enough time to relax. It results in the reduction of the 'mechanical strength' of the grain boundaries in the UFG alumina.

The decrease in the fracture toughness coefficient of alumina in the range of small grain sizes (Fig. 3a) is another indirect evidence in favor of this suggestion. The presence of defects and small pores at the grain boundaries promotes the arising of microcracks when indenting. In this connection, the decreasing of $K_{IC}$ at $d < 2.1$ μm can also be related to the increased defect density of the ceramic grain boundaries.

**Conclusions**

1. Ceramics with high relative densities were made by SPS from industrial submicron and fine alumina powders. The regime with low heating rate of 10 ºC/min allowing minimizing the probability of cracking of the specimens and ensuring high level of uniformity of the ceramic microstructure was used to obtain the ceramics.

2. The effect of the grain sizes on the mechanical properties (hardness, fracture toughness, dynamic ultimate strength) of the fine-grained alumina has been investigated. The increasing of the grain sizes was shown to result in a monotonous decrease in the hardness and not to affect the minimal fracture toughness coefficient essentially. The dependence of the hardness on the grain size can be described by the formula $H_v - d^{-1/2}$ with a good precision. The dependence of the dynamic ultimate strength on the mean grain size has a non-monotonous character with a maximum corresponding to the grain size of 1.9-2.1 μm. The origin of the reduction of the dynamic strength of the fine-grained alumina with the mean grain size less than 2 μm is probably the reduction of the relative density and increased defect density at the ceramic grain boundaries.

3. The specimens of the fine-grained alumina with the grain sizes 1.9-2.1 μm have a high relative density (97.98%), good hardness (18.4 GPa), fracture toughness (2.7 MPa·m$^{1/2}$) and dynamic compression ultimate strength (1060 MPa) at the strain rate $10^3$ s$^{-1}$.

**Acknowledgements**

The present study was supported by Russian Science Foundation (Grant #20-73-10113).

The investigations of the powders by Transmission Electron Microscopy were carried out at the NUST MISIS (supported by Ministry of Science and Higher Education of the Russian Federation, Project #075-15-2021-696).

**References**

1. Shevchenko V.Ya., Barinov S.M. Tekhnicheskaya keramika [Technical Ceramics], Moscow: Nauka Publ., 1993, 192 p. (in Russ.)

2. Boldin M.S., Berendeev N.N., Melehin N.V., Popov A.A., Nokhrin A.V., Chuvil'deev V.N. Review of ballistic performance of alumina: Comparison of alumina with silicon carbide and boron carbide. *Ceramics International*. 2021. V. 47. Iss. 18. P. 25201-25213.

3. Evans A.G., Langdon T.G. Structural Ceramics. Pergamon Press. 1976, 308 p.

4. Bragov A.M., Chuvil'deev V.N., Melekhin N.V., Boldin M.S., Balandin V.V., Nokhrin A.V., Popov A.A. Experimental study of dynamic strength of aluminum oxide based fine-grained ceramics obtained by Spark Plasma Sintering. *Journal of Applied Mechanics and Technical Physics*. 2020, v.61. iss.3, p.494-500.

5. Murray N.H., Bourne N.K., Rozenberg Z., Field J.E. The spall strength of alumina ceramics. *Journal of Applied Physics*. 1998. V. 84. Iss. 2. P. 734-738.

6. Belenky A., Rittel D. Static and dynamic flexural strength of 99.5% alumina: Relation to porosity. *Mechanics of Materials.* 2012. V. 48. P. 43-55.

7. Krell A., Blank P. Grain size dependence of hardness in dense submicrometer alumina. *Journal of the American Ceramic Society*. 1995. V. 78. Iss. 4. P. 1118-1120.

8. Muche D.N.F., Drazin J.W., Mardinly J., Dey S., Castro R.H.R. Colossal grain boundary strengthening in ultrafine nanocrystalline oxide. *Materials Letters*. 2017. V. 186. P. 298-300.

9. Tokita M. Progress of Spark Plasma Sintering (SPS) Method, Systems, Ceramics Application and Industrialization. *Ceramics*. 2021. V. 4. Iss. 2. P. 160-198.

10. Munir Z. A.; Anselmi-Tamburini U.; Ohyanagi M. The effect of electric field and pressure on the synthesis and consolidation of materials: A review of the spark plasma sintering method. *Journal of Materials Science*. 2006. V. 41. Iss. 3. P. 763-777.

11. Olevsky E., Dudina D., Field-Assisted Sintering. Springer, Cham. 2018, 425 p.

12. Tokita M. Spark Plasma Sintering (SPS) Method, Systems, and Applications (Chapter 11.2.3). In Handbook of Advanced Ceramics (Second Ed.). Ed. Shigeyuki Somiya, Academic Press. 2013, p.1149-1177.

13. Orlova A.I. Crystalline phosphates for HLW immobilization – composition, structure, properties and production of ceramics. Spark Plasma Sintering as a promising sintering technology. *Journal of Nuclear Materials*. 2022. V. 559. 153407.

14. Chuvil'deev V.N., Boldin M.S., Dyatlova Ya.G., Rumyantsec V.I., Ordan'yan S.S. Comparative study of hot pressing and Spark Plasma Sintering of $Al_2O_3/ZrO_2/Ti(C,N)$ powders. *Russian Journal of Inorganic Chemistry*, 2015, v.60, No.8, p.987-993.

15. Bragov A.M., Grushevsky G.M., Lomunov A.K. Use of the Kolsky method for confined test of soft soils. *Experimental Mechanics*. 1996. V. 36. P. 237-242.

16. Staehler J.M., Predebon W.W., Pletka B.J., Lankford J. Testing of high-strength ceramics with the split Hopkinson pressure bar. *Journal of the American Ceramic Society.* 1993. V. 76. Iss. 2. P. 536-538.

17. Wang Z., Li P. Characterisation of dynamic behaviour of alumina ceramics: evaluation of stress uniformity. *AIP Advances*. 2015. V. 5. 107224.

18. Hu Z.-Y., Zhang Z.-H., Cheng X.-W., Wang F.-C., Zhang Y.-F., Li S.-L. A review of multi-physical fields induced phenomena and effects in spark plasma sintering: Fundamentals and applications. *Materials and Design*. 2020. V. 191. 108662.

19. Anselmi-Tamburini U., Gennari S., Garay J.E., Munir Z.A. Fundamental investigations on the spark plasma sintering/synthesis process: II. Modeling of current and temperature distribution. Materials Science and Engineering A. 2005. V. 394. Iss. 1-2. P. 139-148.

20. Rahaman M.N. Ceramic Processing and Sintering. New York: Marcel Dekker Inc. 2003, 876 p.

21. Bradt R.C. et. al. Fracture Mechanics of Ceramics. Part: Indentation Size Effect on the Hardness of Zirconia Polycrystals. Springer. 2005. 636 p

22. Ryou H., Drazin J.W., Wahl K.J., Qadri S.B., Gorzkowski E.P., Feigelson B.N., Wollmershauser J.A. Below the Hall-Petch limit in nanocrystalline ceramics. *ACS Nano*. 2018. V. 12. Iss. 4. P. 3083-3094.

23. Garshiv A.P., Gropyanov V.M., Zaytsev G.P., Semenov S.S. Keramika dlya mashinostroeniya [Ceramics for Mechanical Engineering]. Moscow: Nauchtekhlitizdat Publ., 2003. 384 p. (in Russ.)

24. Dillon S.J., Harmer M.P. Multiple grain boundary transitions in ceramics: A case of study of alumina // Acta Materialia. 2007. V. 55. Iss. 15. P. 5247-5254.

25. Wynblatt P., Rohrer G.S., Papillon F. Grain boundary segregation in oxide ceramics. Journal of the European Ceramic Society. 2003. V. 23. Iss. 15. P. 2841-2848.

26. Mazitov A.B., Oganov A.R. Grain boundaries in minerals: Atomic structure, phase transitions, and effect on strength of polycrystals // Zapiski Rossiiskogo Mineralogicheskogo Obshchestva. 2021. V. 150. Iss. 5. P. 92-102.

27. Betekhtin V.I., Kadomtsev A.G., Kipyatkova A.Yu., Glezer A.M. Excess free volume and mechanical properties of amorphous alloys // Physics of the Solid State, 1998, v.40, iss.1, p.74-78.

28. Chuvil'deev V.N. Nonequilibrium grain boundaries in metals. Theory and applications. Moscow: Fizmatlit, 2004. 304 p. (in Russ.)